\documentclass{article}
\usepackage{ijcai20}

\usepackage{times}

\usepackage{soul}
\usepackage{url}
\usepackage[hidelinks]{hyperref}
\usepackage[utf8]{inputenc}
\usepackage[small]{caption}
\usepackage{graphicx}
\usepackage{amsmath}
\usepackage{booktabs}
\usepackage{graphicx}
\graphicspath{ {./} }
\usepackage{ijcai20}
\usepackage{algorithm}
\usepackage{algorithmic}
\usepackage{tabularx}
\usepackage{graphicx}
\usepackage{adjustbox}
\usepackage{times}
\usepackage{soul}

\usepackage{soul}
\usepackage{url}
\usepackage[hidelinks]{hyperref}
\usepackage[utf8]{inputenc}
\usepackage{amsfonts}
\usepackage{graphicx}
\usepackage{amsmath}
\usepackage{booktabs}
\usepackage{multirow}
\usepackage{tabularx}
\usepackage{booktabs}
\usepackage{siunitx}
\usepackage{caption}
\usepackage{nicematrix}
\usepackage{enumitem}
\usepackage{booktabs}

\title{Effect of Homomorphic Encryption on the Performance of Training Federated Learning Generative Adversarial Networks}

\author{
Ignjat Pejic$^1$, Rui Wang$^2$, and Kaitai Liang$^2$
\affiliations
$^{1,2}$EEMCS, Delft University of Technology, The Netherlands\\
\emails
$^1$i.pejic@student.tudelft.nl,
$^2$\{r.wang-8, kaitai.liang\}@tudelft.nl
}

\begin{document}

\maketitle

\begin{abstract}
A Generative Adversarial Network (GAN) is a deep-learning generative model in the field of Machine Learning (ML) that involves training two Neural Networks (NN) using a sizable data set. In certain fields, such as medicine, the training data may be hospital patient records that are stored across different hospitals. The classic centralized approach would involve sending the data to a centralized server where the model would be trained. However, that would involve breaching the privacy and confidentiality of the patients and their data, which would be unacceptable. Therefore, Federated Learning (FL), an ML technique that trains ML models in a distributed setting without data ever leaving the host device, would be a better alternative to the centralized option. In this ML technique, only parameters and certain metadata would be communicated. In spite of that, there still exist attacks that can infer user data using the parameters and metadata.  A fully privacy-preserving solution involves homomorphically encrypting (HE) the data communicated. This paper will focus on the performance loss of training an FL-GAN with three different types of Homomorphic Encryption: Partial Homomorphic Encryption (PHE), Somewhat Homomorphic Encryption (SHE), and Fully Homomorphic Encryption (FHE). We will also test the performance loss of Multi-Party Computations (MPC), as it has homomorphic properties. The performances will be compared to the performance of training an FL-GAN without encryption as well. Our experiments show that the more complex the encryption method is, the longer it takes, with the extra time taken for HE is quite significant in comparison to the base case of FL.\newline
\end{abstract}

\section{Introduction}
 Machine learning (ML) is a subset of artificial intelligence (AI) that allows models to learn from the data provided to them. One such model is a Generative Adversarial Network (GAN) \cite{one}, which has positively impacted the field of generating fake data that appears to be real. To accurately train the GAN, a considerable amount of data may need to be used, and that data is not always centralized. Therefore to obtain the data, users need to willingly share their data, which is not always the case due to privacy concerns. To avoid such issues, Federated Learning (FL) \cite{two} is a better training technique for GANs. FL is a machine learning model that allows GANs to be trained in a collaborative and distributed fashion, where the data does not have to be shared or communicated outside of the model training facility. The information shared will only consist of the parameters of the locally trained models and metadata, which is a huge improvement from the initial approach. Unfortunately, according to \cite{three}, there is still a risk that malicious users or the main server may be able to infer the data from the parameters and metadata. The malicious user may recover data samples by solving for the optimal pair of inputs and outputs to match the parameters communicated to the main server \cite{leak}. To address this issue, one privacy preserving implementation involves using Homomorphic Encryption (HE) on any data that is to be communicated \cite{three}. This will enable us to train any model that requires sensitive data more accurately as we will be able to obtain more data to train them.
 
 To illustrate the importance of using HE with FL-GANs, we will provide an example. If we wanted to study different medical images, we could use HE with FL-GANs, to train a Generator that is able to produce realistic fake images. We can then create numerous instances of certain types of diagnoses, and be able to further improve our doctors' abilities to study and understand the medical images. That can save human lives, and may not be possible if patients would be unwilling to share their data for fear of privacy invasion.

 While using HE on FL-GANs solves the issue of privacy, there is a significant performance loss. This is where our research topic comes in. The topic is as follows:
 \begin{itemize}
     \item {Implement a Federated Learning Generative Adversarial Network with Homomorphic Encryption, and study the effect of HE on the performance (and accuracy) of training the model.}
 \end{itemize}
 
 We will implement the FL-GAN and study the performance with three different types of HE: Partial Homomorphic Encryption (PHE), Somewhat Homomorphic Encryption (SHE), and Fully Homomorphic Encryption (FHE) \cite{gentry}. We will also use a tool implementing Secure Multiparty Computation (MPC) \cite{mpc}, that has both additive and multiplicative homomorphic properties. Each of these types of encryption (and MPC) will be explained in further detail in the section 2. This paper will not involve proving whether or not the encryption schemes ensures privacy, as it has already been proven that they do \cite{four}.  

 This paper will be organized as follows. Section 2 will contain the background information, where GANs, FL,HE, and MPC will be explained and discussed in further detail. Then, section 3 will go over our System Model, describing the threat model and FL algorithm. Moving on, section 4 will elaborate on our research approach and implementation, where the experiment and tools used will be described. We will go over the implementation of the topics discussed in section 2 and how they fit together to get our desired result. After that, section 5 will go over the results that our experiment has obtained on the performance loss with HE and MPC.  Moving forward, in section 6 we will discuss our results and go over them while also explaining why anything is missing. We will mention any unexpected results, and try to elaborate on the possible reasons for them. In section 7, we will go over the possible future work and research that should be done to increase our understanding of the topic. It will be based on our ideas developed through researching our topic. Lastly, in section 8 we will conclude our research by providing a summary of what was done.

\section{Background Knowledge}
The coming subsections will go over the background knowledge that we deemed essential to understand all parts of the report. We will discuss the GAN, FL, HE, and MPC.
\subsection{Generative Adversarial Network}
 A GAN is a generative machine learning model consisting of two Neural Networks that compete against each other with the goal of being able to generate realistic fake data \cite{five}. The two models in question are Discriminators and Generators. A Discriminator is, at its basic form, a classifier used to distinguish between real and fake data. On the other hand, a Generator is a model, that when provided with a random input (random noise), generates fake data. The process of training a GAN starts with training the Discriminator for a certain number of epochs (number of iterations of data set) on a labelled data set, followed by training the Generator on the Discriminator for also a number of epochs. The training is done using the minimax equation, stated below:
\begin{equation} 
 \min_{G}\max_{D}\mathbb{E}_{x\sim p_{\text{data}}(x)}[\log{D(x)}] +  \mathbb{E}_{z\sim p_{\text{z}}(z)}[1 - \log{D(G(z))}]
\end{equation}

 What the above equation states is that the Discriminator tries to maximize the likelihood of determining the correct label (real or fake) for both the training examples and for examples generated by the Generator. On the other hand, the Generator tries to minimize the probability that the Discriminator can correctly classify its generated images. 
 
 This whole process gets repeated multiple times. The training process is ideally completed when the Discriminator can predict whether generated data is real with a probability of 50\%. As that is practically difficult to achieve, the usual way of training a GAN involves repeatedly training the Generator and Discriminator for a certain number of epochs, and then measuring the accuracy of the generated images.  

\subsection{Federated Learning}
The way a machine learning model is trained can have a huge impact on the privacy of the model. Initially, most ML models were trained in a centralized setting, meaning all of the data had to be sent by the data owners to a main server, where it would be used to train the model. However, in a scenario that involves hospital patient data, for instance, the data owners may be unwilling to share the data due to privacy concerns. In order to maintain data confidentiality, a technique called Federated Learning was developed \cite{seven}.

While there do exist different forms of Federated Learning, the one we will use in this research paper will be a centralized FL using the fed-avg algorithm \cite{eight}. This involves one main central server and multiple client nodes. Each of the client nodes will have its own ML model with randomly initialized model parameters, where it will be trained using the local data. As the training process takes place, the model parameters will be updated for each. Once the clients have trained their models, each client will send its model’s parameters to the centralized server. The server aggregates the parameters, and sends them back to the individual client nodes. The training process then repeats until the models are adequately trained.

An example of an FL-GAN can be seen in figure 1.
\begin{figure}[t]
\includegraphics[width=0.5\textwidth ]{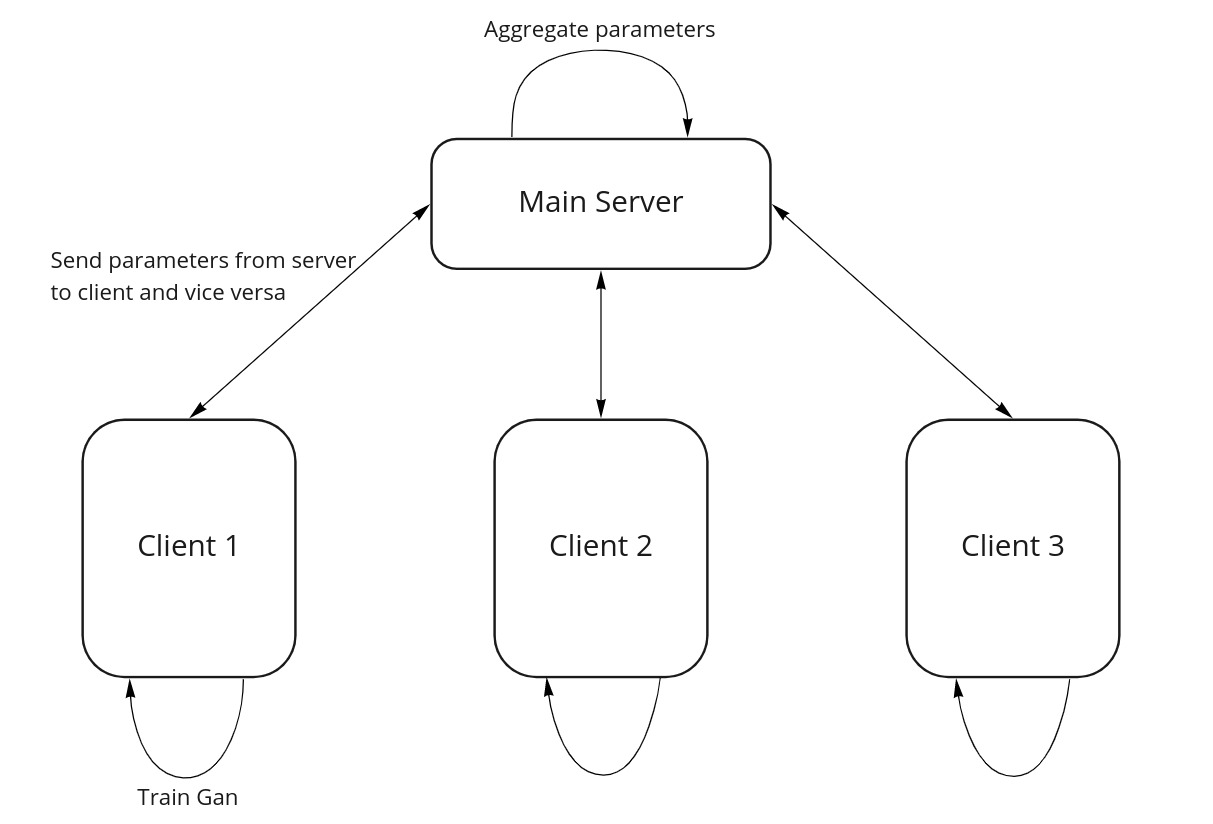}
\centering
\caption{Visualization of FL-GAN}
\end{figure}

\subsection{Homomorphic Encryption}
HE is a technique used to protect data confidentiality of users \cite{nine}. While standard encryption (non HE) is used to make the data unreadable, it does not allow for any computation to be performed on the data. In order to do any computations, the data would first have to be decrypted, which is a privacy risk. On the other hand, data encrypted with HE may be updated with the use of arithmetic computations, which makes it desirable for increasing the privacy in FL and cloud storage/computation.

The formula for basic Homomorphic Encyption is the following:
\begin{equation} c_1 = E_k(x), c_2 = E_k(y)
\end{equation}
\begin{equation} D_k(c_1 + c_2) = x + y
\end{equation}
\begin{equation} D_k(c_1 * c_2) = x * y
\end{equation}
As mentioned beforehand, there are three different types of HE: namely Partial Homomorphic Encryption (PHE), Somewhat Homomorphic Encryption (SHE), and Fully Homomorphic Encryption (FHE).
\begin{itemize}
    \item{PHE allows the user to perform either addition or multiplication (equation 3 or 4) on the encrypted data. In this report, PHE will be assumed to be additive (equation 3).}
    \item{SHE is a more powerful form of encryption, which allows one to perform both additive and multiplicative operations on the ciphertexts (equations 3 and 4). However, there are only a certain amount of operations that can take place. Any operations past this point will not guarantee that our decryption provides the correct result. It is slower than PHE.}    
    \item{FHE is the most powerful form of encryption, but also the slowest. Like the SHE, it allows for both addition and multiplication to be done on encrypted data (equations 3 and 4). The difference is that with FHE there can be an unlimited amount of arithmetic operations that are allowed to take place, using a technique called bootstrapping.}
\end{itemize}

\subsection{Secure Multi Party Computation }
While the main topic of this paper is about the performance loss in training FL-GANs using HE, we also thought it would be interesting to add a tool for comparison that has both additive and multiplicative homomorphic properties (MPC). MPC is a subfield of cryptography, where performing a computation works by distributing it across multiple parties. Each party may only see its own data. This, in combination with the fact that no party further shares its own data, allows for privacy. It has a higher performance than Homomorphic Encryption.

Figure 2 shows an example of MPC.
\begin{figure}[t]
\includegraphics[width=0.5\textwidth ]{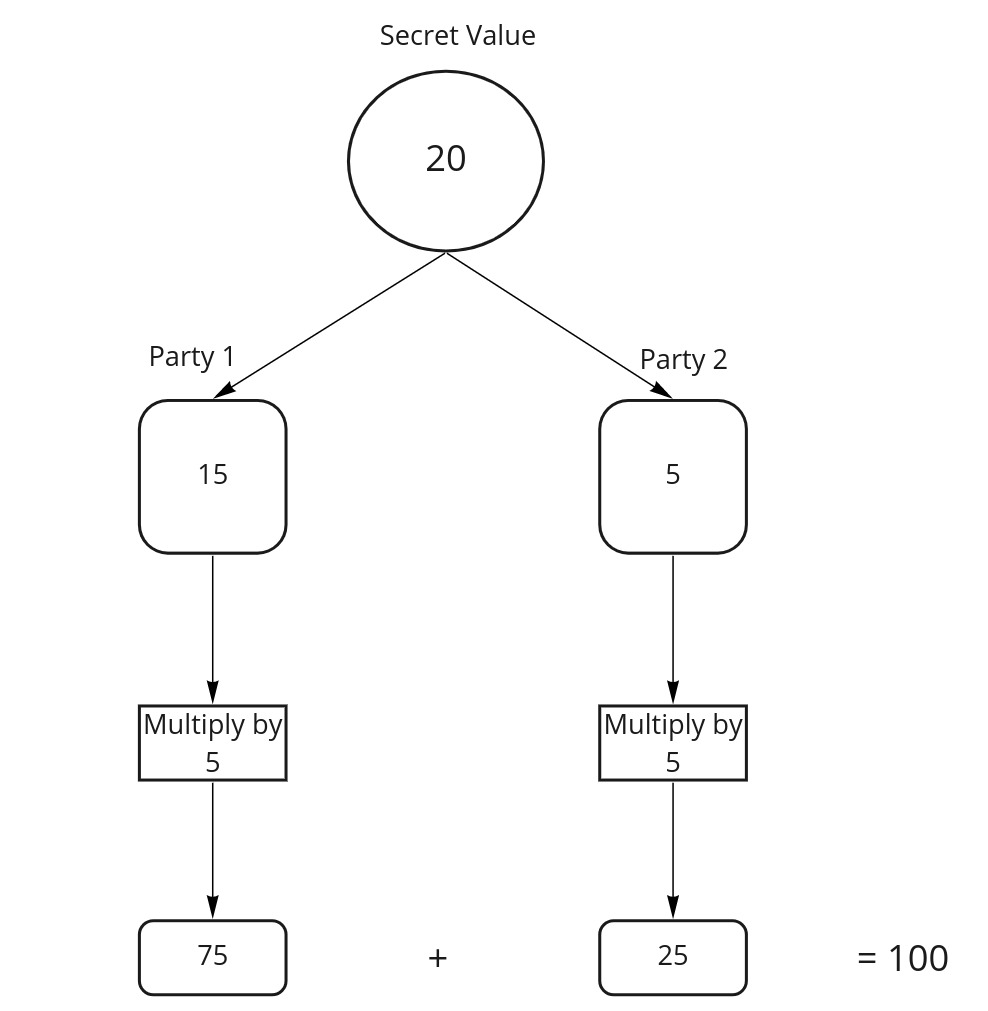}
\centering
\caption{Example of MPC}
\end{figure}

\section{System Model}
This section will go over the proposed implementation of our Federated Learning Generative Adversarial Network with Homomorphic Encryption. We will first go over the threat model, before providing our algorithm of the ML model. 

\subsection{Threat Model}
There are multiple threats that can potentially affect our Federated Learning system, namely the clients that train the GANs, the server that aggregates the parameters received from the clients, and any potential outsiders. We assume our key generation is secure, and its communication to the clients is secure. Moreover, we will consider that both the clients and the server are following the honest-but-curious model, meaning that none will deviate from the algorithm that exists, but will try and learn as much as possible from any received information. When it comes to the potential outsiders, we assume that they may intercept any message except the key generation and its communication. 

\begin{table}[]
\caption{Notation Table}
\begin{adjustbox}{width = 0.5\textwidth}
\begin{NiceTabular}{cc}
\hline
\textbf{Notation}        & \textbf{Definition}             \\ \hline
$n$                      & Number of clients               \\ \hline
$pd$                     & Parameters of Discriminator     \\ \hline
$pg$                     & Parameters of Generator         \\ \hline
$dd$                      & Dataset                         \\ \hline
$s$                      & Sum of parameters               \\ \hline
$rp$                     & Parameters received from server \\ \hline
$\emptyset$ & Empty set                       \\ \hline
\end{NiceTabular}
\end{adjustbox}
\end{table}

\subsection{Experiment Algorithm}
Our algorithm will consist of a centralized server, that will send each partition of the main dataset to its corrsponding client node. The client node will then train its individual GAN on the received dataset, before encrypting the parameters and sending them back to the server. The server will then use the fed-avg algorithm \cite{fedavg} to aggregate the results. This will, in the simplest terms, involve averaging the parameters of all clients, before sending them back to each client. The clients will then decrypt the received parameters, update the model, and keep training the model for as many epochs as necessary. 

For the client side algorithm, refer to Algorithm 1, while for the server side use Algorithm 2. Make sure to use table 1 for all the notations to understand the algorithms.

At the beginning of the process, each public and private key is generated by the key generator and then sent to each of the clients. As mentioned previously, we assume that the communication of the keys is fully secure. The key generators no longer participate in any part of our machine learning model and receive no data whatsoever. 

\begin{algorithm}
\caption{Algorithm 1: GAN training on client}
\begin{algorithmic} 
\REQUIRE $pub\_key$, $priv\_key$, $dataset$
\STATE $d \leftarrow discriminator()$
\STATE $g \leftarrow generator()$
\WHILE{$dd \neq \emptyset$}
\STATE \texttt{train($d$, $g$)}
\ENDWHILE
\FOR{$p$ in $pg$}
\STATE $p \leftarrow p/n$
\STATE $p \leftarrow Enc(p)$
\STATE \texttt{send encrypted parameter to server}
\STATE \texttt{receive aggregated parameter from server}
\STATE $p \leftarrow Dec(p)$
\ENDFOR
\FOR{$p$ in $pd$}
\STATE $p \leftarrow p/n$
\STATE $p \leftarrow Enc(p)$
\STATE \texttt{send encrypted parameter to server}
\STATE \texttt{receive aggregated parameter from server}
\STATE $p \leftarrow Dec(p)$
\ENDFOR    
\end{algorithmic}
\end{algorithm}

\begin{algorithm}
\caption{Algorithm 2: Server}
\begin{algorithmic} 
\STATE $dd \leftarrow import dataset$
\FOR{$c$ in $n$}
\STATE $partition\_c \leftarrow dd/n$
\STATE \texttt{send $partition\_c$ to $client\_c$}
\ENDFOR

\STATE \texttt{Wait for clients to train GAN}

\FOR{$p$ in $pg$}
\STATE $s \leftarrow \emptyset$
\FOR{$c$ in $number\_of\_clients$}
\STATE $s \leftarrow s + rp$
\ENDFOR
\STATE \texttt{send parameters back to clients}
\ENDFOR

\FOR{$p$ in $pd$}
\STATE $s \leftarrow \emptyset$
\FOR{$c$ in $n$}
\STATE $s \leftarrow s + rp$
\ENDFOR
\STATE \texttt{send parameters back to clients}
\ENDFOR

\end{algorithmic}
\end{algorithm}

\section{Research Approach and Experiment Setup}
This section will discuss the overall approach to begin answering the research question, alongside all of the considerations that needed to be taken into account. This section will dive more into the implementation details in comparison to the algorithm of section 3.

\subsection{Research Approach}
Initially, after going through the existing research on FL, GANs, and HE, a number of observations were made. There exists a lot of research on each topic individually, on the GANs trained with FL, and on the effect of HE on FL. However, we have been unable to find research encompassing all three fields simultaneously. 

What was also unfortunate is that the majority of the research papers did not have the code base used to conduct their experiments shared publicly, making it challenging to recreate their results or further improve on their own. Moreover, for the repositories that were publicly shared, a lot of documentation was either missing or incorrect, leading to the same issues as stated above.

Due to the reasons stated in the previous paragraph, it became clear that we would need to integrate the three components ourselves. Given the time constraints of the projects, we decided that the best approach was to look at industry standard tools used for secure machine learning, as we expected them to have the best documentation and support of use. What we have not been able to implement will be found in section 8 as the future work.
\subsection{Experiment}
As stated in the Research Approach subsection, we decided to use the most used industry standard tools for our integration of GANs, FL, and HE. We will go over the implementation process, starting with unsuccessful attempts before moving on to our approach that turned out successful.

To begin with, there are two main open source repositories used for Machine Learning, namely PyTorch \cite{torch}, developed by Meta/Facebook, and TensorFlow \cite{tf}, developed by Google. Using these softwares, it is possible to create most of the standard ML models. There exist many implementations of GANs in both, allowing us to easily construct and train our own model. When it comes to training those models using FL, there exist a number of implementations including preexisting frameworks: PySyft \cite{syft} for PyTorch and TensorFlowFederated for TensorFlow \cite{tff}. The final step would be to be able to use HE alongside such models. This is where the main issue appeared.

Both PyTorch and TensorFlow at one point had their own frameworks for homomorphic encryption: PySyft for PyTorch again and TF-Encrypted for TensorFlow. However, as of May 2022, both of the frameworks are unusable for this project. Starting with PySyft, the software has recently been and still is being updated. While the software is being updated, the documentation is still the same. While there do exist many examples of HE with PyTorch in the past, they do not work with the new version of PySyft. Moving on to TF-Encrypted, the repository has not been updated recently. Therefore, it is currently incompatible with the newest version of TensorFlow. As a consequence of neither framework allowing for direct HE, we have decided to work on creating an implementation using PyTorch, as we found it easier to use and modify to our advantage. 

We started out by implementing a FL-GAN using the Message Passing Interface (MPI) \cite{mpi}, which was initially created by a group member. It allows for parallel computing and has been tried and tested. For our experiment, we decided to initially use four nodes, consisting of one server node and three client nodes. The Neural Networks of the GAN, the Discriminator and Generator, are trained on each client node. The data used for training the GAN is the CIFAR-10 dataset, which consists of 60000 32x32 color images of 10 categories, namely airplanes, cars, birds, cats, deer, dogs, frogs, horses, ships, and trucks. The 60000 images are split even among the categories. 

After the models are trained, their parameters are sent to the main server node via a data structure called a tensor. The tensor has a similar structure to a n-dimensional matrix, but is specific to PyTorch. The default datatype of each element is a float32. At this stage, the server aggregates the data using an algorithm called fed-avg. This technique involves taking the mean of each parameter. After the aggregation takes place, the data is sent back to each client for the training process to continue. This process is repeated for as many epochs as desired.

At this point we had an implementation of the FL-GAN, but we were still missing HE. In order to include it, the next step was to encrypt all parameters on the client side before they were communicated with the server. Where possible in our implementation, we tried to avoid giving our server the encryption keys.

Unfortunately, two issues were present. Firstly, due to the implementation of MPI, it was not trivial to send the keys from the key generator to the clients. Namely, certain methods of implementation required encryption contexts. Sending those contexts using MPI is not possible, as that is one of the things that MPI forbids to be sent and received. Moreover, certain forms of encryption slowed down the main process by a large amount, making it unfeasible to run on our machine. 

Due to the concerns above, we have instead decided to approximate the extra time taken to train the FL-GAN with HE. To achieve this, we have recorded the time taken to generate all of the necessary encryption keys. Afterwards, we recorded the time taken to both encrypt and decrypt each value. Using these calculations and our knowledge of the total number of parameters involved in training the GAN, we approximated the time taken for all the necessary encryptions and decryptions. Moreover, where necessary, we measured the difference in size between plaintexts and encrypted ciphertexts, and used that to approximate the extra time taken for communication between the clients and the server. The main algorithm used for our entire implementation can be found in section 3.

\section{Results}
After implementing our model, we have recorded/approximated the performance loss of training FL-GANs with HE. The first subsection will discuss the results of running the model without any HE, and each following subsection will present the additional performance loss for a specific type of HE or MPC. Some terms used in this section include $t$, the number of tensors, $p$, the total parameter data found in all of the tensors for both the Generator and Discriminator, $c$, the number of clients, and $e$, the number of (federated) epochs. In our experiments, $t$ is 13 for the Generator and 11 for Discriminator, $p$ is 6342272, $c$ is 3, and $e$ is 50. 

All the tests were implemented in Python 3.7, and were performed on a machine with a Intel (R) Core (TM) i7-8750H processor with 2.20GHz. Due to the way MPI works, the number of cores used depends on the number of clients. There is an additional core for the main server and one for a key generator whenever necessary.

\subsection{Base Case - FL-GAN With No HE}
As mentioned in sections 3 and 4, we have run our implementation of the FL-GAN using three clients and a central server, aggregating the parameters using an algorithm called fed-avg. We have run the federated algorithm for a total of 50 epochs, with each of the three clients using a third of the CIFAR-10 dataset for training the GAN. Their individual dataset remains constant for the entirety of the training process. 

After our implementation has finished running, we have collected our results. We have only run the process three times, due to the length of time necessary. The average training process took a total of 16 hours, and we measured the accuracy using Binary Cross Entropy Loss (BCELoss). BCELoss is the loss observed in binary classification. It increases for the Discriminator when it incorrectly classifies a real or generated image, and increases for the Generator if the Discriminator believes that the generated image is fake.

\begin{figure}[t]
\includegraphics[width=0.5\textwidth ]{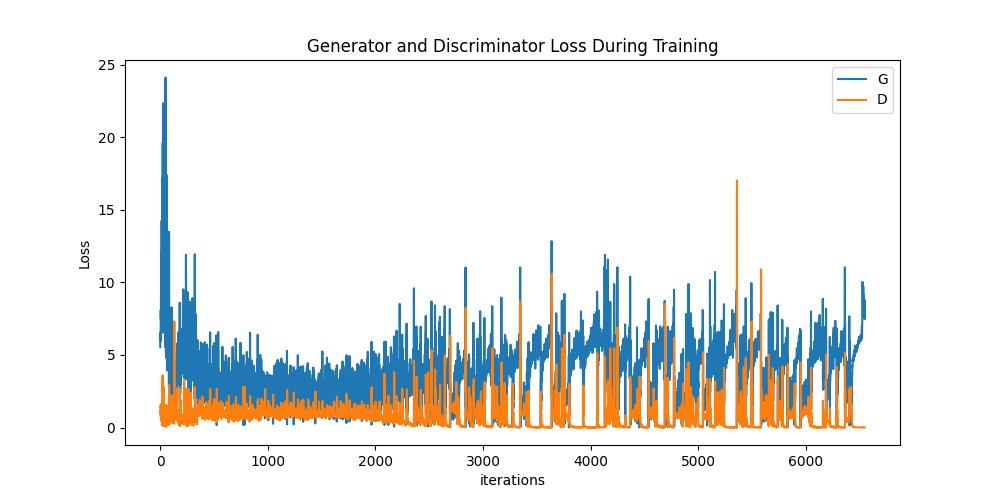}
\centering
\caption{BCELoss for Generator and Discriminator as the number of training iterations increases}
\end{figure}

\subsection{FL-GAN with PHE}
In this subsection, we have used an algorithm for Partial Homomorphic Encryption, namely the Paillier Cryptosystem (PC) \cite{paillier}. This algorithm enables us to do add any ciphertexts or multiply them by a constant value. The tool we used as our PC is \cite{pp}, as our research showed that this library is used and referenced the most in the Python language. 

In our implementation, the worst case time complexity for adding the PC is $\mathcal{O}(p)$. Ideally, we would be able to encrypt each tensor, which is expected to take less time and be more optimized. Unfortunately, given that those tools are being updated, we had to iterate over each parameter in each tensor and encrypt each of those values.

We have decided to compare the performance loss of adding the PC using 4 different keys, namely: 64 bits, 128 bits, 256 bits, and 512 bits. To approximate the extra time needed, we have recorded the time taken to first generate the keys. This was less than 0.1 seconds, making it negligible. Then, the time taken to encrypt and decrypt one value was recorded. For both of these recordings, an average time was taken. The average of a minimum of 10000 runs was done for each recording. We then had to multiply the time taken for one encryption and decryption by $p$, to get the total encryption and decryption time for one client for one training epoch. Lastly, we multiplied that by $c$ and then by $e$. We have also recorded the size of the ciphertext for each key length. It is worthwhile mentioning that while the ciphertexts were of different size, the Paillier object that was stored and communicated was always 64 bits. Due to some implementation issues, we decided that instead of communicating tensors, we would first convert them to numpy arrays, encrypt the values, and then communicate them for aggregation. The time taken for converting between a tensor and a numpy array was negligible (less than 2 seconds total). Moreover, the aggregation also did not add up to any significant time (also less than 2 seconds). 

All of the other results can be found in table 2. They will include the total time lost depending on the key size. As our results only contain the performance loss for 3 clients and 50 epochs, equation 5 will state the general formula for the total time lost.
\begin{equation} total\;time\;lost = p*(encryption + decryption\; time) * c * e
\end{equation}
Figure 4 visually presents the total time lost for each of the four key sizes. Lastly, as the PC has no losses with encryption and decryption, there is no accuracy loss in training the GAN in comparison to the base case.

\begin{table}[]
\caption{Results for Paillier Cryptosystem for 3 clients and 50 epochs}
\centering
\begin{adjustbox}{width = 0.5\textwidth}
\begin{NiceTabular}{lllll}
\hline
\textbf{K.S. \tabularnote{ Key Size} } & \textbf{T.F.O.E. (s) \tabularnote{ Time for one encryption}} & \textbf{T.F.O.D. (s) \tabularnote{Time for one decryption}} & \textbf{T.A.T.P.C.P.E. (s) \tabularnote{Total additional time per client per epoch}} & \textbf{T.A.T.T. (s) \tabularnote{Total additional time taken (3 clients,50 epoch)}} \\ \hline
64                & 0.000105                           & 1.612e-5                           & 771                                        & 115650                                          \\ \hline
128               & 0.0005040                          & 0.00018                            & 4311                                       & 646650                                           \\ \hline
256               & 0.00134                            & 0.00058                            & 12177                                      & 1826574                                          \\ \hline
512               & 0.00671                            & 0.00023                            & 44015                                      & 6602305                                          \\ \hline
\end{NiceTabular}
\end{adjustbox}

\end{table}
\begin{figure}[t]
\includegraphics[width=0.5\textwidth ]{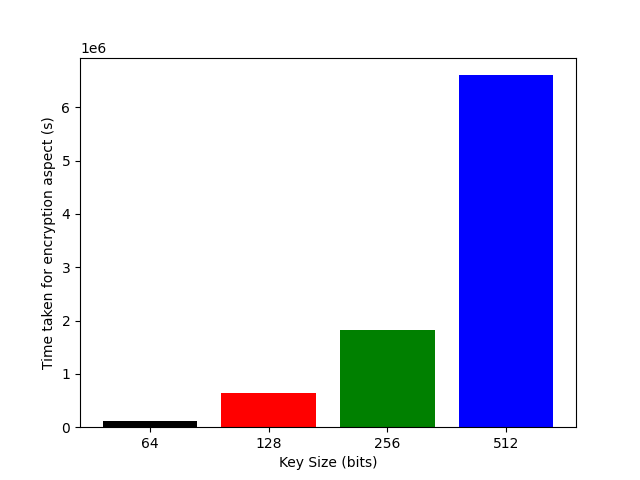}
\centering
\caption{Total time lost due to PHE depending on key size}
\end{figure}

\subsection{FL-GAN with SHE}
When it comes to SHE, we have come across two main schemes. The first is the Brakerski-Fan-Vercauteren (BFV) scheme \cite{bfv}, which is used when working with integers. The second scheme is the Cheon-Kim-Kim-Song (CKKS) scheme \cite{ckks}, which allows for encryption and decryption of real numbers. Both of these schemes are layered, meaning that we there is a limit on the number of arithmetic operations we can make on the ciphertext before it becomes undecipherable. As our tensor stores values which are float32, we have decided to go with the CKKS scheme. 

As with all other SHE implementations, we can do both addition and multiplication between cipher texts. Not only that, as CKKS is based on Learning With Error (LWE), it is quantum secure. We were fortunate to find an implementation of CKKS that works with our PyTorch implementation: TenSEAL \cite{tenseal}. TenSEAL is built on top of Microsoft SEAL, the industry standard to both SHE and FHE. 

When it comes to predicting the total time loss, we have decided to go with two different approaches. The first is similar to what we did with the PC, where we encrypted every parameter of every tensor. This also has a time complexity of $\mathcal{O}(p)$. However, due to the fact that TenSEAL is compatible with PyTorch, we were able to encrypt tensors as a whole, improving our time complexity to $\mathcal{O}(t)$. The first approach was done to have a suitable comparison with the PC. We are only able to use a 128 bit key with TenSEAL, so that was our key strength for both of the approaches. The performance based results for both of the approaches can be found in table 3. 

For approach one, we can also use equation 5 to predict the total loss for any number of clients and epochs. For approach 2, we can refer to equation 7.

\begin{equation} tt = total\;tensor\;encryption\;and\;decryption\;time \end{equation}
\begin{equation}total\;time\;lost = tt * c * e
\end{equation}

\begin{table}[]
\caption{Results for SHE for 3 clients and 50 epochs}
\centering
\begin{adjustbox}{width = 0.5\textwidth}
\begin{NiceTabular}{lllllll}
\hline
\textbf{A \tabularnote{Approach}} & \textbf{E.T.P.I.$P$. (s)\tabularnote{Encryption time p in $p$}} & \textbf{E.T.T.I.$T$. (s) \tabularnote{Encryption time per t in $t$}} & \textbf{D.T.P.I.$P$. (s) \tabularnote{Decryption time p in $p$}} & \textbf{D.T.T.I.$T$. (s) \tabularnote{Decryption time t in $t$}} & \textbf{T.P.C.P.E. (s) \tabularnote{Time per client per epoch }} & \textbf{TT (s) \tabularnote{Total time}} \\ \hline
1        & 0.00397                      & N/A                              & 0.00113                          & N/A                              & 32322                         & 4848408        \\ \hline
2        & N/A                          & 0.00487                          & N/A                              & 0.140 - 0.377                    & 14.25                         & 2137           \\ \hline
\end{NiceTabular}
\end{adjustbox}

\end{table}

Unfortunately we do not only have a time based performance loss. CKKS supports approximate arithmetic calculations over real numbers, meaning that our operations will have some losses in accuracy, affecting our entire model. The reasoning is that at a certain point in the CKKS encryption scheme, we need to round our transformed ciphertext. This rounding will cause a loss. More on this can be found here \cite{ckksloss}. 

As mentioned in section 4.2, MPI is unable to communicate everything, including contexts. As TenSEAL uses contexts, we are unable to communicate both the keys and ciphertexts, not allowing us to present any performance metric.

\subsection{FL-GAN with FHE}
Regarding FHE, we do not have any results due to time constraints. We were planning to run the same CKKS scheme that we had mentioned in the previous subsection, but would add bootstrapping to the system. We again would have had to use only the CKKS scheme, due to the data type. We predict that this would have taken a lot of extra time. This will further be mentioned in the discussions sections 6.2 and 8.

\subsection{FL-GAN with MPC}
Similarly to TenSEAL, we were able to find a software repository that used Secure Multi Party Computation that was compatible with PyTorch. The tool in question is Crypten \cite{crypten}. Using Crypten, we were able to encrypt whole tensors at a time, making the time complexity $\mathcal{O}(t)$. While the previous subsections had an approach encrypting every parameter of every tensor, we decided that this was unnecessary for MPC as it is not HE, making an exact comparison between two methods of lower priority. The results for this approach can be found in table 4. As with the second approach of SHE, we can use equation 7 to estimate the time loss for any number of clients and epochs.

\begin{table}[]
\caption{Results for MPC for 3 clients and 50 epochs}
\begin{adjustbox}{width = 0.5\textwidth}
\begin{NiceTabular}{llll}
\hline
\textbf{TET (s)} \tabularnote{Tensor Encryption Time} & \textbf{TDT (s)} \tabularnote{Tensor Decryption Time} & \textbf{TTPC (s)} \tabularnote{Total time per client} &
\textbf{TT (s)} \tabularnote{Total time}   
\\ \hline
       0.000184 - 0.0774                           &       0.000153 - 0.03821                            & 0.368                           & 55.2 \\ \hline
\end{NiceTabular}
\end{adjustbox}

\end{table}

This approach also has accuracy losses, with each parameter in each tensor being off by 2.2798291e-05. This difference, over 6342272 parameters will add up to an accuracy loss.

\section{Discussion}
This section will discuss two topics. We will first go over our results, and then mention what results are missing and we will explain the reasons.

\subsection{Result Analysis}
Overall, most of the results that we got were as expected. When it comes to the additional time increase, we expected Multi Party Computations to be the quickest, followed by Partial Homomorphic Encryption, and lastly Somewhat Homomorphic Encryption. 

The base case scenario of running our FL-GAN on the CIFAR-10 dataset with 3 clients ran in 16 hours. The images generated were similar to \cite{similarloss}, a repository use to train a normal GAN on the CIFAR10 dataset. The images produced show progress as the training process continues for the most part, but as Figure 3 shows at a certain point the BCELoss stops decreasing, and then starts variating inconsistently. This was unexpected, but we believe the reason could be due to the fact that we over trained the Discriminator. We can see in the figure that in many instance were the Discriminator loss is near 0, the Generator loss increases. As the Discriminator gets better and better trained, it can become more challenging for the Generator to be trained properly, as almost anything not perfect may be rejected by the Discriminator.

When it comes to the PC, the stronger the security was the longer it took to encrypt all the necessary parameters. The total time increase between the 64 bit key and the 512 bit key was from just above 32 hours to just under 1834 hours. This time difference is extreme, as can be seen on Figure 4, and demonstrates how important it is to balance security and performance. They key we will use as a standard for comparison with other results is the 128 bit key, which would take roughly 180 hours to run. It would take more than 11 times the training process of classical FL-GAN to just complete the encryption and decryption with the PC. 

Moving on to the CKKS scheme for SHE, we see that there is a huge difference between the two approaches (encrypting every parameter in every tensor and encrypting every tensor as a whole). The first approach took roughly 1347 hours compared to 36 minutes of the second approach. We expected the second approach to be quicker, but this difference in efficiency is shocking. It makes us think of the results we would have gotten if we were able to do the second approach on the Paillier Cryptosystem. This will be expanded in section 8. However, when it comes to comparing approach 1 of CKKS to the PC, we can see that CKKS is more than 7 times less efficient. As CKKS is a lot more complex, and given the fact that we can do both addition and multiplication with it, this does not come as a surprise. There also is a loss in accuracy, which we have unfortunately been unable to measure.  

The last results we got were with regards to MPC. As predicted, this provided the least time overhead, only 55 seconds. It was faster than any other differential privacy tool we used in this experiment by far. While there is no reason not to use MPC given such a low performance loss, there does appear to be a slight loss in accuracy (value of 2.2798291e-05 per parameter). This adds up over all the parameters, clients, and epochs, and would most definitely increase our BCELoss. Overall, the security properties are lower for MPC than for Homomorphic Encryption, but the low overhead may make such an option enticing. 

Figure 5 shows the difference in extra time overhead for the PC (128 bits), SHE (128 bits) encryption per parameter per tensor, SHE (128 bits) encryption time per tensor, and MPC.

\begin{figure}[t]
\includegraphics[width=0.5\textwidth ]{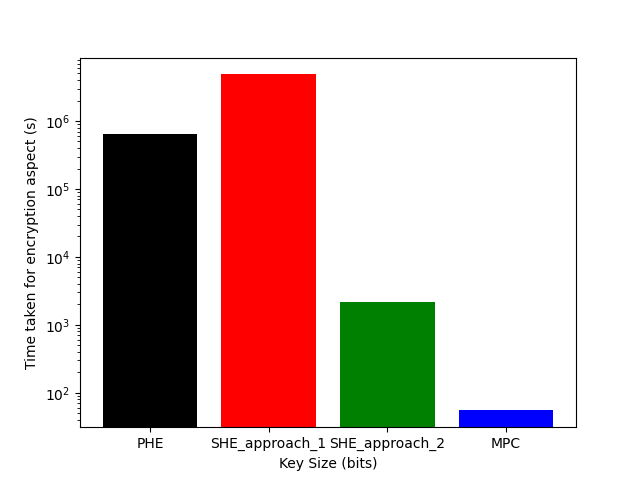}
\centering
\caption{Total time lost between PHE, SHE, and MPC}
\end{figure}

\subsection{Missing Results and Improvements}
To begin, we are missing any results when it comes to Fully Homomorphic Encryption. While we used the CKKS scheme, which has the bootstrapping option, we never used it. The reason is our federated algorithm did not contain any multiplicative operations. Moreover, we only had 3 clients, making bootstrapping unnecessary. It would also have required a lot more time, making it difficult to fit within our time constraints.

Another improvement that could have been made was using the DelftBlue cluster \cite{db}. The miscalculation here was missing out on the fact that every MPI node runs on one CPU, thereby no mater how many CPUs total we had there was no performance boost. Had we used a different implementation, we could have been able to run some of the implementations rather than predicting their results.

\section{Future Work}
This section will go over what we believe should be then next when it comes to our research topic.

The first thing we believe should be done is trying out the effect of HE on FL-GANs using different implementations. There do exist a lot of other forms of repositories that allow for us to make an FL model, such as Fate \cite{fate}, and they all may have their own tools that are optimized for implementing HE. It would be interesting to compare amongst the different forms of implementation.

Also, given that most HE tools (such as Microsoft SEAL) are written in c and c++, which are both programming languages that have better optimizations than python, one could implement the whole project in those languages. On the other hand, most ML models are implemented with Python, so choosing different languages would mean less online support and less flexibility to alter the models that do exist.

When it comes to our implementation using PyTorch, it would be ideal to test it with PySyft when the documentation is updated. That would give us the options of encrypting tensors as a whole using the Paillier Cryptosystem, which would give individuals and corporations more knowledge on how big of an impact HE would actually have on the models.

Moreover, as mentioned throughout this paper our FL was implemented through MPI. An industry standard FL-GAN would be implemented in a different way, where the communication costs between the clients and the server would probably take considerable extra time, which we currently do not take into account. Also, with a real implementation we can get the performance metrics of the accuracy of different HE schemes. In addition to that, there is a high likelihood that the system would have considerably more clients than three, which definitely needs to be further looked into and experimented. Using different federated algorithms would also be useful, as they may have a different amount of arithmetic operations on the ciphertexts. This could affect a leveled CKKS scheme, especially when combined with more clients.

Finally, it would be crucial to also experiment on FHE. As this is the strongest form of HE, finding out how much slower it is than SHE and PHE would be significant. As mentioned in the previous paragraph, this should also be done for different federated algorithms.  

With all of the above mentioned sections, it should be noted that more time should be spent researching the accuracy loss of training the model, as the results in our experiment were more focused on performance. 

\section{Conclusion}
When it comes to the training of Machine Learning models, nothing trumps the importance of privacy and data confidentiality of individuals. While Federated Learning is a significant improvement as only parameters of a model leave the model training location, it is still not ideal as bad actors are in certain situations able to reverse engineer the parameters to obtain the original data. One of the solutions to this issue is using Homomorphic Encryption, an encryption technique allowing us to perform computations on ciphertexts. This way, we only communicate encrypted parameters, which ensures that no unauthorized party has any access to the initial data used to train the ML model.

The objective of this paper is to find the performance loss of training a Federated Learning Generative Adversarial Network using Homomorphic Encryption. From the three types that exist (Partial, Somewhat, and Fully HE), we were able to predict the performance loss of Partial and Somewhat HE (for our implementation). We have used the Paillier Cryptosystem for PHE and the Cheon-Kim-Kim-Song for SHE. All of our results are for a FL structure consisting of three clients, but where applicable we have also added a formula that can be used to estimate the performance loss for a n-client system. Moreover, we have also predicted the performance loss of Multi Party Computation, as it has homomorphic properties. Our results support the fact that as our encryption system gets stronger, the performance loss is higher, making the decision of balancing security and performance a difficult but nevertheless vital issue for the developers. 

As we implemented and worked on our research topic, we have come across many different FL implementation and HE libraries that are optimized for those implementations. That opens the possibility for a future research on those industry standard tools, and comparing the results amongst themselves. It will be interesting to see the effect on efficiency as the Homomorphic Encryption techniques gets more optimized.

\bibliographystyle{plain}
\bibliography{references}

\end{document}